\documentclass[10pt]{amsart}
\allowdisplaybreaks[4]
\usepackage{amssymb,color}
\usepackage{amsmath,amssymb,graphicx}
\usepackage{accents}
\usepackage{bbm}
\usepackage{lipsum}
\usepackage{array}
\usepackage{multirow}

\usepackage{tabularx,ragged2e,booktabs,caption}
\usepackage{here}
\usepackage[toc,page]{appendix}

\usepackage[colorlinks=true, citecolor=blue, linkcolor=blue, urlcolor=blue ]{hyperref}

\usepackage{amsthm}
\usepackage{subcaption}

\numberwithin{equation}{section}
\numberwithin{figure}{section}
\numberwithin{table}{section}
\allowdisplaybreaks[4]

\definecolor{c20}{rgb}{0.,0,0.}
\definecolor{c30}{rgb}{1,0.5,0.5}
\definecolor{c40}{rgb}{0,.7,0}
\definecolor{c50}{rgb}{1,0,1}

\def\TE#1{\textcolor{c30}{#1}}
\def\eEe#1{\textcolor{c50}{#1}}
\def\eEe#1{#1}
\def\TE#1{#1}

\def\mE#1{\textcolor{c30}{#1}}

\def\mE#1{#1}

\newtheorem{theo}{Theorem}[section]
\newtheorem{sat}[theo]{Proposition}
\newtheorem{de}[theo]{Definition}
\newtheorem{lem}[theo]{Lemma}
\newtheorem{exxa}[theo]{Example}

\newtheorem{korr}[theo]{Corollary}

\newtheorem{remarks}[theo]{Remarks}

\newcommand{\kb}[1]{\boldsymbol{#1}}
\newcommand{\vk}[1]{\kb{#1}}

\def\E#1{\mathbb{E}\left \{#1 \right\}}

\newcommand{\pk}[1]{\mathbb{P} \left\{ #1 \right\} }

\newcommand{\R}{\mathbb{R}}

\newcommand{\ldot}{,\ldots,}

\newcommand{\BQN}{\begin{eqnarray}}
\newcommand{\EQN}{\end{eqnarray}}

\newcommand{\BQNY}{\begin{eqnarray*}}
\newcommand{\EQNY}{\end{eqnarray*}}

\newcommand{\BS}{\begin{sat}}
\newcommand{\ES}{\end{sat}}
\newcommand{\BT}{\begin{theo}}
\newcommand{\ET}{\end{theo}}
\newcommand{\BK}{\begin{korr}}
\newcommand{\EK}{\end{korr}}

\newcommand{\BD}{\begin{de}}
\newcommand{\ED}{\end{de}}

\newcommand{\BIT}{\begin{itemize}}
\newcommand{\EIT}{\end{itemize}}
\newcommand{\BDI}{\begin{description}}
\newcommand{\EDI}{\end{description}}

\newcommand{\BRM}{\begin{remarks}}
\newcommand{\ERM}{\end{remarks}}

\newcommand{\BTH}{\begin{theo}}
\newcommand{\ETH}{\end{theo}}
\newcommand{\BPR}{\begin{sat}}
\newcommand{\EPR}{\end{sat}}

\newcommand{\BEX}{\begin{exxa}}
\newcommand{\EEX}{\end{exxa}}

\newcommand{\BC}{\begin{cases}}
\newcommand{\EC}{\end{cases}}
\newcommand{\COM}[1]{}
\newcommand{\BL}{\begin{lem}}
\newcommand{\EL}{\end{lem}}

\def\ldot{, \ldots,}
\def\vv{ \mathcal{V}_{@\mathcal{R}} }
\def\PY{P_i^*}
\def\PIY{P_i+ \tau_i}
\def\piY{\pi_i(\PIY)}

\begin{document}

\title[
Tariff \& Premium optimisation ]{Some Mathematical Aspects of Price Optimisation}

\author{Yizhou Bai}
\address{Yizhou Bai, School of Mathematical Sciences, Nankai University, PR China 
and Department of Actuarial Science, University of Lausanne, UNIL-Dorigny 1015 Lausanne, Switzerland}

\author{Enkelejd Hashorva}
\address{Enkelejd Hashorva, Department of Actuarial Science, University of Lausanne, UNIL-Dorigny 1015 Lausanne, Switzerland}

\author{Gildas Ratovomirija}
\address{Gildas Ratovomirija, Department of Actuarial Science, University of Lausanne, UNIL-Dorigny 1015 Lausanne, Switzerland, 
and 	Vaudoise Assurances, Place de Milan CP 120, 1001 Lausanne,         Switzerland}

\author{Maissa Tamraz}
\address{Maissa Tamraz, Department of Actuarial Science, University of Lausanne, UNIL-Dorigny 1015 Lausanne, Switzerland}

\bigskip

\date{\today}
 \maketitle

\bigskip
\begin{quote}
{\bf Abstract}: Calculation of an optimal tariff is a principal challenge for pricing actuaries.  
In this contribution we are concerned with the renewal insurance business discussing various mathematical aspects of calculation of an optimal renewal tariff. 
Our motivation comes from two important actuarial tasks, namely a) construction   of an  optimal renewal tariff 
subject to business and technical constraints, and b) determination of an optimal allocation of certain premium loadings. 
We consider both continuous and discrete optimisation and then present several algorithmic sub-optimal solutions. Additionally, we explore some simulation techniques. Several illustrative examples show both the complexity and the importance of the optimisation approach.  

\end{quote}

{\bf Key Words}: market tariff; optimal tariff;  optimal price; price elasticity; non-life insurance; non-convex optimisation; quadratic programming; sequential 
quadratic programming; mixed discrete non-linear programming; constraints; renewal business\\



\def\atR{$@ \mathcal{R} \ $}
\def\atRR{$@ \mathcal{R}$}

	\section{Introduction}
Commonly, insurance contracts are priced based on a tariff, here referred to as the {\it market tariff}.  In mathematical terms such a tariff is a 
function say $f: \R^d \to [m,M]$ where $m,M$ are the minimal and the maximal premiums. For instance, 
a motor third party liability (MTPL)  {\it market tariff} of key insurance market players in Switzerland has  $d> 15$. 
Typically, the function $f$ is neither linear nor a product of simple functions.\\ 
In non-life insurance, many  insurance companies use different $f$ for new business and renewal business. 
There are statistical and marketing reasons behind this \TE{practice}. In this paper we are primarily concerned with non-life 
renewal business. Yet, some findings are of importance for general pricing of insurance and other non-insurance products.
We shall discuss two important actuarial tasks and present various mathematical aspects of   relevance for 
pricing actuaries. \\
 
\underline{\bf Practical actuarial task T1}: {\it Given that a portfolio of $N$ policyholders is priced under a given {market tariff}  $f$, 
determine an optimal {\it market tariff} $f^*$ that will be applied in the next portfolio renewal.}   \\

Typically, actuarial textbooks are  concerned with the calculation of the pure premium, which is determined by 
applying different statistical and actuarial methods to historical portfolio data, see e.g., \cite{Embetal1997,Rolski,MR1794582,MR3244191}. The tariff that determines the pure premium of a given insurance contract will be here referred to as the {\it pure risk  tariff}. In mathematical terms this is a function say 
$g: \R^{d_1} \to [m_1, M_1]$ with $d_1 \ge 1$. 

In the actuarial practice, pure premiums are loaded,  for instance for large claims,  provisions, 
direct expenses and other costs (overheads, profit, etc.). 

Actuarial mathematics explains various approaches to load  premiums; in practice very commonly a linear loading is applied. 
We shall refer to the function that is utilised for the calculation of the premium of an insurance coverage based on the costs related to that coverage 
 as {\it actuarial tariff};  write $g_A:\R^{d_2} \to [m_2, M_2]$ for that function.\\ 
 
\underline{\bf Practical actuarial task T2}: {\it Given a pure risk tariff $g$, construct 
 an optimal actuarial tariff $g_A$ that \TE{includes} various premium loadings.} 

Since by definition there is no unique optimal {\it actuarial tariff}, the calculations leading to it can be performed 
depending on the resources of pricing and implementation team. \\
To this end, let  us briefly mention an instance which motivates {\bf T2}: Suppose for simplicity that the portfolio 
in question consists of two groups of policyholders A and B. In group A there are $n_A$ policyholders and in group $B$ there are already $n_B$ policyholders. All the contracts are to be renewed at the next 1st January. 
The pricing actuary calculates the {\it actuarial tariff} which shows that for group A, the yearly premium to be paid from each policyholder is 2'000 CHF and for group B, 500 CHF. For this portfolio, overhead expenses (not directly allocated to an insurance policy) are calculated (estimated) to be X CHF for the next insurance period (one year in this case). 
The amount $X$ can be distributed to $N=n_A+n_B$ policyholders in different ways, for instance each policyholder will have to pay $X/(n_A+n_B)$ of those expenses. Another alternative approach could be to calculate it as a fix percentage of the pure premiums. 
The principal challenge for pricing actuaries is that the policyholders already are in the portfolio and know their current premiums.

At renewal (abbreviated as \atR in the following)   given that the risk does not change, if the new offered premium is different from the current one, the policyholder can cancel the contract. Another reason for cancelling the policy could also be the competition 
in the insurance market. Consequently, the solution of both {\bf T1} and {\bf T2} needs to take into account the probability of cancellation of the policies at the point of renewal. Both {\bf T1} and {\bf T2} are in 
general very difficult to solve. A simpler problem in renewal pricing is the following:\\ 

\underline{\bf Practical  actuarial task T3}: {\it Modify for any $i\le N$ the premium  $P_i$ of the $i$th policyholder \atR  by a fixed percentage, say \mE{$\delta_i$ with $\delta_i \in \Delta_i=\{ 0\%, 5\%\}$} so that the new set of premiums
 $$P_i^*= P_i+ \tau_i, \quad  1 \le i\le N \mE{\text{ with } \tau_i= P_i\delta_i}$$
  are optimal. Moreover, determine the new market tariff  $f^*$ which  yields $P_i^*$'s. }
 
Indeed, the actuarial task described in {\bf T3} is very common in actuarial practice, if the actual performance of the portfolio is not as expected, and a premium increase is to be applied at the next renewal.

There are several difficulties related to the solutions of tasks {\bf T1-T3}. In practice the {\it market tariff} is very complex for key insurance coverages such as motor or household insurance. A typical $f$ used in  practice is as follows (consider only two arguments for simplicity)

\def\bqny#1{ { \begin{eqnarray*} #1 \end{eqnarray*}}}
\def\bqn#1{ { \begin{eqnarray} #1 \end{eqnarray}}}
\bqn{ \label{exf}
f(x,y)= \min\Bigl(M_0, \max( e^{a x+ bx},m_0 + m_1 x+ m_2 y) \Bigr).  
}
Even if we know the  $P_i^*$'s that solve {\bf T3}, when the structure of $f$ (and also of $f^*$) is fixed say as in \eqref{exf}, then the existence of an optimal $f^*$ that gives exactly $P_i^*$'s is in general not guaranteed. 
Note that due to technical reasons, the actuaries can change the coefficients that determine $f$, say $a,b$ and so on, but the structure of the tariff, i.e., the form of $f$ in \eqref{exf} is in general fixed when preparing a new renewal tariff  due to huge implementation 
costs.

  The main goal of this contribution is to discuss various mathematical aspects that lead to optimal solutions 
  of the actuarial tasks {\bf T1-T3}. Further we analyse eventual implementations of our optimisation problems for renewal business. 
Optimisation problems related to new business are much more involved and will therefore be treated in a forthcoming contribution. \\
To this end, we observe that in the last 10 years many insurance companies in Europe have already used price optimisation techniques (mostly through consultancy companies).  So far in the literature, there is no precise mathematical description  of the optimisation problems solved in such applications. 
\TE{Very recent contributions focus on the issues of price optimisation, mainly from the ethical and regulation points of view, see 
\cite{PROA,PROB}.} \eEe{It is important to note that optimality issues in insurance and reinsurance business, not directly related to the problems treated in this contribution,  have been discussed in various context, see \cite{AsimitRiskTransfer,MR3130463,MR3430410, Opt1, Opt2, Opt3, Opt4} and the references therein.} 
\\  

Brief organisation of the rest of the paper: \mE{Section 2 describes the different optimisation settings from the insurer's point of view. In Section 3, we provide partial solutions for problem \textbf{T1}. Section 4 describes the different algorithms used to solve the optimisation problems followed by some insurance applications to the motor line of business  presented in Section 5.}


\section{Objective functions and Business Constraints} 
\subsection{Theoretical Settings}
For simplicity, and without loss of generality,  we shall assume that the renewal time is fixed for all $i=1, \ldots, N$ policyholders already insured in the portfolio with the $i$th policyholder paying $P_i$ for the current insurance period. Each policyholder can be insured for different insurance periods. Without loss of generality, 
 we shall suppose that at renewal each insurance contract \TE{has the option to be renewed for say one year}, with a renewal premium of $P_i+ \tau_i $. \\
  Suppose that the cancellation probability for the \TE{$i$th} contract is 
a function of $P_i$. At renewal, by changing the premium, the cancellation probability  will depend on the premium change, say $\tau_i$ and the initial premium $P_i$. Therefore we shall assume that this probability is given by 
\bqn{\label{PI}
\mE{\piY= \Psi_i(P_i, \delta_i),  \text{ with } \tau_i=P_i\delta_i},}
where $\Psi_i$ is a strictly positive monotone function depending eventually on $i$. This is a common assumption in logistic regression, where $\Psi_i$ is the inverse of the logit function (called also expit), or $\Psi_i$ is a univariate distribution function.\\
In order to consider the cancellation probabilities in the tariff and premium optimisation tasks, the actuary needs to know/determine $\pi_i(P_i + \tau_i)$ 
for any \mE{$\delta_i \in \Delta_i$}, where $\Delta_i$ is the range of possible changes of premium with $0 \in \Delta_i$. 
Estimation of $\pi_i$'s is difficult, and can be handled for instance using logistic regression, see Section \ref{probrenwMb} below for more details. 

In practice, depending on the market position and the strategy of the company, different objective functions can be used for the determination of an optimal {\it actuarial tariff} or {\it market tariff}.  We discuss below two important objective functions: 
\def\NN{ N_{@ \mathcal{R}}}
\BIT
\item[{\bf O1})] \underline{Maximise the future expected premium volume \atRR}: \\
\TE{In our model}, the current premium volume for the portfolio in question is $V=\sum_{i=1}^N P_i$, whereas the premium volume in case of complete renewal 
 is $\sum_{i=1}^N \PY= \sum_{i=1}^N (P_i+ \tau_i).$ Since not all policies might renew, let us denote by $\NN$ the number of policies which  will be renewed. Since we can treat each contract as an independent risk, then 
 $$\NN = \sum_{i=1}^N I_i,$$
  with $I_1 \ldot I_N$ independent Bernoulli random variables with 
 $$\pk{I_i=1}= \piY , \quad 1 \le i\le N.$$
Clearly, the expected percentage of the portfolio to renew is given by 
 \bqn{
 \theta(\tau_1 \ldot \tau_N)= \frac{\E{\NN}}{N}= \sum_{i=1}^N \frac{\E{I_i}}{N} = 
\frac 1 {N} \sum_{i=1}^N \piY  .
 }
The premium volume \atR (which is random) will be denoted by $\vv$. It is simply given by 
$$ \vv:=\sum_{i=1}^N I_i (P_i+ \tau_i).$$  
Consequently, the objective function is given by (set below $\vk{\tau}=(\tau_1 \ldot \tau_N)$)
\bqn{\label{qvol} q_{vol} (\vk{\tau} ):= \E{\vv}= \sum_{i=1}^N (\PIY) \E{I_i} =
\sum_{i=1}^N (\PIY) \piY .
}   
Note that $P_1 \ldot P_N$ are known, therefore the optimisation will be performed with respect to $\tau_i$'s only. 
\item[{\bf O1'})] \underline{Minimise the variance of \TE{$\vv$}}: If the variance of $\vv$ is large, the whole renewal process can be ruined. Therefore along {\bf O1} the minimisation of the variance of  $\vv$  is important. In this model we have 

\bqn{ \label{vvv}
q_{var}(\vk{\tau}):=Var( \vv )&=& 
   \sum_{i=1}^N(\PIY) \piY  [1- \piY ] .  
}

\item[{\bf O2})] \underline{Maximise the expected premium difference \atRR}:  The premium difference for each policyholder in our notation is $\tau_i$ and thus at renewal we have $\sum_{i=1}^N I_i \tau_i$. The expectation of this random variable is simply 
\bqn{\label{qdif}
 q_{dif}(\vk{\tau})= \E{\sum_{i=1}^N I_i \tau_i}= \sum_{i=1}^N \tau_i \piY .
 }
\EIT
It is not difficult to formulate other objective functions, for instance  related to the classical ruin probability, Parisian ruin, or future solvency and market position of the insurance company. Moreover, the objective functions can be formulated over multiple insurance periods. 

\COM{
the objective of the optimisation is typically one of the following 
(for a given time-period, say 1 year),
\BIT
\item O1) Maximise the future expected profit; 
\item O2) Maximise the future expected premium volume; 
\item O3) Minimise the ruin probability.
\EIT
In actuarial practice, for obvious reasons, O2) is the preferred optimisation target. 
}

\TE{Due to the nature of insurance business, there are several constraints that should be taken into account, \eEe{see for instance \cite{AsimitConstraints} and 
	the references therein}. Typically, the most important business 
contraints relate to the strategy of the company and the concrete insurance market.} We formulate few important constraints below: 

\BIT 
\item[{\bf C1})] Expected retention level \atR should be bounded from below: Although the profit and the volume of  premiums at renewal are  important, 
all insurance companies are interested in keeping  most of the policyholders in their portfolio. Therefore there is commonly a lower bound on the expected retention level $\ell \in [0.7, 1]$ at renewal. For instance $\ell= 90\%$ \TE{means that the expected percentage of customers that 
will not renew their contrat should not exceed } 10\%. 
In mathematical terms, this \TE{is formulated as}  
\bqn{\label{retentionC1}
\theta_{rlevel}(\vk{\tau})= \frac{\E{N^*}}{N} \ge \ell.
}
\item[{\bf C2})]  A simple constraint is to require that the renewal premiums $P_i^*$'s are not too different from the "old" ones, i.e., 
\bqn{\label{AB}
\frac{\tau_i }{P_i} \in [a,b], \quad \tau_i \in [A,B], \quad 1 \le i\le N 
}
for instance \mE {$a=-5\%$ and $b=10\%$} and $A= -50, B=300$. 
\EIT
Several other constraints including those related to reputational risk, 
decrease of provision level for tied-agents, and loss of loyal customers can be formulated similarly and will therefore not be treated in detail.

\def\vvd{ \mathcal{V}^c_{ @ \mathcal{R}} }

\subsection{Practical Settings} In insurance practice the cost of optimisation itself (actuarial and other resources) 
needs to be also taken into account. Additionally, since the total volume of premium at renewal is 
large, an optimal renewal tariff is of interest if it produces a significant improvement to the current tariff.
Therefore, for practical implementations, we need to redefine the objective  functions. For a given positive  constant $c$, say $c=1'000$, we redefine \eqref{qvol} as 
\bqn{\label{qvolc} q_{vol}^c (\tau_1 \ldot \tau_N):= c \Bigl \lfloor  \E{\vv}/c  \Bigl \rfloor = 
c \Bigl \lfloor\sum_{i=1}^N (\PIY) \piY /c  \Bigl \rfloor ,
}   
where $\lfloor x \rfloor$ denotes the largest integer smaller than $x$. 
Similarly, we redefine \eqref{vvv} as 
\bqn{ \label{vvv}
c \Bigl \lfloor Var( \vv )/ c  \Bigl \rfloor &=& 
c \Bigl \lfloor   \sum_{i=1}^N(\PIY) \piY  [1- \piY ] / c  \Bigl \rfloor.  
}
Finally, \eqref{qdif} can be written as 
\bqn{\label{qdifC}
 q^c_{dif}(\tau_1 \ldot \tau_N)= c \Bigl \lfloor  \sum_{i=1}^N \tau_i \piY / c  \Bigl \rfloor.  
 }
For implementation purposes and due to business constrains, $\tau_i$'s can be assumed to be certain given numbers. 
Therefore a modification of \eqref{AB} can be as follows 
\bqn{\label{ABC}
\delta_i:= \frac{\tau_i }{P_i} \in [a,b] \cap (c_1^{-1} \mathbb{Z}) , \quad \tau_i \in [A,B] \cap (\TE{c_1} \mathbb{Z}), \quad 1 \le i\le N, 
}
where $c_1>0$, for instance $c_1=100$.

Such modifications of both objective functions and constraints show that for practical implementation, there is no unique optimal solution of the optimisation problem of interest.

\BRM \TE{i) In this contribution we are not directly concerned with distributional channels. For example,  if two policies say the $i$th and the $k$th ones are renewed through different distributional channels, then perhaps different contraints are to be applied to each of those policies. Additionally, the cancellation probabilities could be different, even in the case where both policyholders have the same risk profile. Therefore, in order to allow for different distributional channels, we only need to adjust the constraints and assume an appropriate cancellation pattern.\\
ii) From the practical point of view, $\Psi_i$'s are estimated by using for instance logistic regression. At random, customers are offered higher/lower premiums than their $P_i$'s at renewal, i.e., $\tau_i$'s are chosen randomly with respect to some prescribed distribution function. An application of the logistic regression to the data obtained (renewal/non renewal) explains the cancellation (or renewal) probability in terms of risk factors as well as other predictors (social status, etc.)
In an insurance market dominated by tied-agents this approach is quite difficult to apply. \\
iii) Different policyholders can renew for different periods. This case is included in our assumptions above. \\
iv) Most tariffs, like say a MTPL one,  consist of hundreds of coefficients (typically more than 400). 
Due to a dominating product-structure, advanced tariffs contains many individual cells, say 200'000 in average. However, most of these tariff cells are empty. For instance, it is quite rare that a Ferrari is insured for a TPL risk by a 
90 years old lady, living in a very small village. With this in mind, typically, the relevant number $N$ in practical optimisation problems does not exceed  50'000.  Our algorithms and simulation methods work fairly well for such $N$.
}
\ERM
\COM{
Whereas the list of constraints is huge, 
It is not particularly difficult to write the constraints and the objective functions in a proper mathematical setting. 
In this contribution we shall present a general mathematical model where the optimisation problems can be easily formulated. 
 The difficulty lies in the concrete  mathematical 
 various  based on  explain the describe the calculation of the premiums    roughly speaking is In order for 
The rest of the paper is organised as follows. In section 2, we define the assumptions taken into account in our study to model the price elasticity. Section 3 presents the optimisation problem and the methodology of the algorithm used to solve it.  A numerical example is established using different methods for solving the optimisation problem. The application of the optimisation problem based on real insurance data is described in Section 4. A conclusion and perspectives are given in Section 5.

In this paper, our aim is to provide an understanding of price optimisation in non-life insurance for renewal business. We would like to present a new method for solving the optimisation problem focusing on the price elasticity of the policyholders. In this respect, an objective function is defined based on the objective of the insurer, profit or volume objective subject to constraints on the volume portfolio or the insurer profit respectively. We suppose that the pure risk premium are actuarially fair and are known. We use the logistic  model to address the probability of renewal of an insured which is a function of the variation of premium over one year time horizon. Our goal is to find the optimal variation of premium at the individual policyholder level by solving the optimisation problem above using several algorithms.\\

}

\section{Solutions for  {\bf T1-T3}} 
The chief difficulty when dealing with the actuarial tasks {\bf T1-T3} lies on the complexity of $\Psi_i$'s since these functions are:  \\
a) in general not known, \\
b) difficult to estimate if past data are partially available,\\
c) even when these functions are known, the constraints {\bf C1-C2} and the objective functions {\bf O1,O1',O2} 
are in general not convex. We discuss next a partial solution for {\bf T1}: \\

\underline{\bf Problem T1a}:  Given $P_1 \ldot P_N$, determine $\vk{\tau}^*=(\tau_1^* \ldot \tau_N^*)$ such that 
\bqn{  q_{vol}(\vk{\tau}^*) \text{ is maximal}, \quad    q_{var}(\vk{\tau}^*) \text{ is minimal}
 }
under the constraints 
\bqny{
\theta_{rlevel}(\vk{\tau}) \ge \ell,  \quad  \delta_i= \frac{\tau_i}{P_i} \in [a,b], \quad 1\le i \le N. 
}

\underline{\bf Problem T1b}: Determine $f^*$ from $P_1^* \ldot P_N^*$. \\
The solution (an approximate one) of {\bf T1b} can be easily derived. 
Given $P_1^* \ldot P_N^*$, and since the structure of the {\it market tariff} $f$ is known, then $f^* $ can be 
determined (approximately), by running a non-linear regression analysis.\\
Therefore, below we focus on {\bf T1a}.
 Consequently, the main question that we shall discuss here  is how to determine optimal premiums $P_i^*$'s at renewal. 

In insurance practice, the functions $\Psi_i,i\le n$ can be assumed to be piece-wise linear and non-decreasing. 
This assumption is indeed reasonable, since for very small $\tau_i$ or $\delta_i$, the policyholder 
will not be aware of premium changes. The latter assumption can be violated if for instance at renewal the competition modifies 
also their new business premiums. For simplicity, these cases will be excluded in our analysis, and thus we assume that the decision 
for accepting the renewal offer is not influenced by the competitors. \\
We list below some tractable choices for $\Psi_i$'s:
\BIT 
\item[{\bf Ma})]  Suppose that for given known constants $\pi_i, a_i,b_i$  
\bqny{ \mE{\Psi_i(P_i, \delta_i)}= \pi_i (1+ a_i \delta_i+ b_i \delta_i^2),  \quad 1 \le i\le N.
}
In practice, $\pi_i, a_i,b_i$ need to be estimated. Clearly, the case that $b_i'$s are equal to 0 is quite simple and tractable.\\
Note in passing that an extension of the above model is by allowing $a_i$ and $b_i$ to differ depending on  the sign of \mE{$\delta_i$}. 
\item[{\bf Mb})] One choice motivated by the logistic regression model commonly used for estimation of cancellation  probabilities 
is the expit function, i.e., 
\bqny{ 
\mE{\Psi_i(P_i, \delta_i)}= \frac{1}{1+ c_i^{-1} e^{- T_i \delta_i }}, \quad 1 \le i \le N,
}
where $c_i,T_i$'s are known constants (to be estimated in applications).\\
We note that Model {\bf Ma)} can be seen as an approximation of Model {\bf Mb}).
\item[{\bf Mc})] A simple specification is when $\Psi$ is given only for few values of $\delta_i$'s as follows. 
n illustrative data of a policyholder $i$  \mE{are} presented in \ref{table:example_Policy} 
 	\begin{center}
	\begin{tabular}{|c|c|c|c|c|c|c|c|c|c|}
	
	\hline
	$ index $ & 1 & 2 & 3 & 4 & 5 & 6 & 7 & 8 & 9 \\
	
	\hline
	$\delta_{i}$( in $\%$) & -20 \% & -15 \% & -10\% & -5\% & 0\% & 5\% & 10\% & 15\% & 20\% \\
	
	\hline
	$P_{i}(1+ \delta_i)$ & 80 & 85 & 90 & 95 & 100 & 105 & 110 & 115 & 120 \\
	
	\hline
	$\Psi_{i}(P_i, P_i \delta_i)$  & 0.999 & 0.995 & 0.990 & 0.975 & 0.950 & 0.925 & 0.900 & 0.875 & 0.825 \\
	
	\hline
	\end{tabular}
	\captionof{table}{Renewal probabilities as a function of premiums for the $i$th policyholder. }
	\label{table:example_Policy}
\end{center}
\EIT	
The Model {\bf Ma}) is simple and tractable and it can be seen as an approximate model of a more complex one. 
Moreover, it leads to some crucial simplification of the objective functions in question. 
\COM{
 Moreover, it suggests that also instead of the complicated objective functions, we can consider modified ones by using 
 an approximation idea. For instance, assuming that $\delta_i$'s are relatively small, we can use the following approximation 
 \bqn{
 q_{vol}(\vk{\tau}^*)= q_{vol}(\vk{\tau}+ \vk{\delta})\approx  q_{vol}(\vk{\tau})+ \sum_{i=1}^N P_i( 1+ A_i \delta_i+ B_i \delta_i^2).
 }the objective funct
 }
 
 {\bf Toy Model 1}:  We suppose that  $P_1=P_2=  \cdots = P_N$. This leads to a simplification of the objective function.

  {\bf Toy Model 2}: Assume that \mE{$\Psi_i(P,\delta)= \Psi(P,\delta)$} for any $i\le N$, i.e., we assume that all policyholders have the same behavior with respect to cancellation probability.

\COM{
	\section{Notation and Preliminaries}
We consider a portfolios of size $N$ with $P_i^9$ the current premium paid from the $i$th policyholder. 
 the premium paid  The

		\subsection{Assumptions} 
		In our framework, we assume that the overall costs of the insurance business are assumed to be known. Furthermore,  the optimisation is for a one-year time horizon which means that   given the premiums at time $0$ our task is to determine the premium at time $1$. Specifically, for a policyholder $i$
			\BIT 
		\item the premium $P_i^0$ at time $0$ is known,
	\item a list of premiums $P^1 _{i}$, that can be set to the policyholder $i$ at time $1$, is known
		\BQNY
		 	P^1 _{i} =(1+\delta_{i})P^0 _i,
	\EQNY
	where $\delta_{i}$ is the variation  of the premium $P^0 _i$ from time $0$ to time $1$,
	
	\item each value of $P^1 _{i}$ corresponds to a probability $\pi _{i}$ that the policyholder will  renew his policy at time $1$. Similarly to $P^1 _{i}$ , $\pi _{i}$ depends on the variation of premium $\delta_{i}$ and is given by
	\BQNY
		 	\pi_{i} = \pi( \delta_{i})
	\EQNY
	where $\pi$ is a decreasing function.  \\
An illustrative data of a policyholder $i$  is presented in \ref{table:example_Policy} 
 	\begin{center}
	\begin{tabular}{|c|c|c|c|c|c|c|c|c|c|}
	
	\hline
	$j$ & 1 & 2 & 3 & 4 & 5 & 6 & 7 & 8 & 9 \\
	
	\hline
	$\delta_{i}$($\%$) & -20 & -15 & -10 & -5 & 0 & 5 & 10 & 15 & 20 \\
	
	\hline
	$P^1_{i}$ & 80 & 85 & 90 & 95 & 100 & 105 & 110 & 115 & 120 \\
	
	\hline
	$\pi_{i}$  & 0.999 & 0.995 & 0.990 & 0.975 & 0.950 & 0.925 & 0.900 & 0.875 & 0.825 \\
	
	\hline
	\end{tabular}
	\captionof{table}{Probability of renewal and new premiums for a policyholder $i$ with $P^0 _i=100$}\label{table:example_Policy}
\end{center}
		\EIT
		
		\subsection{Demand elasticity model}
	\BQN \label{eq: probRenewal}
					\pi^i( \delta_i)= \frac{1}{1+\frac{e^{-T_i \delta_i}}{c_i}} ,
		\EQN
	 where 
	 $\delta_i$ is the premium variation,
	 $c_i$ and $T_i$ are parameters which depend on the profile of the policyholder.
	 For a given probability of renewal without premium change $\pi^i(0)$ , $c_i$ can be estimated by
	\BQNY
	 			\hat{c}_i= \frac{\pi^i(0)}{1-\pi^i(0)}, 
		\EQNY
		which suggests that $c_i$ determines how the policyholder $i$ behave next year if the insurer does not change the premium. Similarly, the parameter  $T_i$ defines the demand elasticity relative to premium change.
	As an example, we assume that the probability of renewal for a policyholder without premium change is $0.95$. 
	
\section{Optimal premiums} 
\subsection{Objectives and constraints}
Every mathematical optimisation consists in maximising an objective function subject to some constraints.
In our case the objective function is related to the objective of the insurer. Thus, we assume that the insurer would like to maximise the total profit $S$ of its portfolio over a one year time horizon. Since the overall costs of the insurer are assumed to be fixed, our task consists in maximising the expected  earned premium of the insurer. Hence, given $n$ policyholders, the corresponding objective function is described as follows
		\BQNY
 \mathbb{E}(S)=\sum_ {i=1}^{n}(1 + \delta_{i})P^0 _i \pi( \delta_{i}).
 		\EQNY
Regarding the constraints, they are also linked to the insurance business, principally on the volume of the portfolio and/or on the insurer profit. For example, the insurer would like to keep at least $80 \%$ of the previous volume of its portfolio and/or to earn at least $110\%$ of its previous profit.
 Furthermore, in many countries, insurance companies are subject to government regulations.  For instance, the regulator imposes the insurer to set actuarially fair premium to the policyholders.
Moreover, the pricing strategy is subject to the insurance market structure. Actually, due to  competition in the market, the insurer premiums should be in line with the competitors premiums. Also, we assume that the insurer is subject to some restrictions concerning the minimum  expected number of policyholders $N$ in his portfolio and is given by:
 \BQNY
\sum_ {i=1}^{n} \pi( \delta_{i}) \geqslant N .
 		\EQNY
}

\section{optimisation Algorithm} 
\subsection{O1) Maximise the future expected premium volume \atRR}
In this section, we denote by $t=0$ the present time and by $t=1$ the time at renewal. 
We consider the case where the insurer would like to maximise the future expected premium volume while simultaneously keeping a minimum number of policyholders in his portfolio at $t=1$. Therefore, we denote by $\ell$ the retention level. $ N \ell$ is just the minimum number of policyholders that the insurer would like to keep in his portfolio $@ \mathcal{R}$. \\
In this respect, the optimisation problem can be formulated as follows
\BQN\label{eq:optimprob}
\begin{aligned}
&\underset{\vk{\delta}}{ \text{max }} \sum_ {i=1}^{N} P _i (1 + \delta _{i}) \Psi_i(P_i,\delta_i),\\
&\text{subject to  }  \frac{1}{N}\sum_ {i=1}^{N} \Psi_i(P_i,\delta_i)\geqslant \ell. \\
\end{aligned}
\EQN

\subsubsection{Probability of renewal $\Psi_i$ as in \textbf{Ma)}}
We consider the case where the probability of renewal $\Psi_i$ is of the form $\Psi_i:= \Psi_i(P_i,\delta_i)= \pi_i (1+a_i \delta_i+ b_i \delta_i^2)$ as defined in \textbf{Ma)}.\\
\BIT
\item Setting $b_i=0$, we have
 $$ \Psi_i:=\Psi_i(P_i,\delta_i)= \pi_i (1+a_i \delta_i),$$
where the condition $\Psi_i \in (0,1)$ should hold for all policyholders $i\le N$. Actually, it is satisfied when  $a_i \in (1-\frac{1}{\pi_i},\frac{1}{\pi_i}-1)$ for all $i \leq N$.\\
Since $b_i=0$, then (\ref{eq:optimprob})  is just a quadratic \TE{programming (QP)} problem subject to linear constraints and can be rewritten as follows
\BQN\label{eq:QPstlinear}
\begin{aligned} 
&\underset{\vk{\delta}}{\text{max }} \sum_ {i=1}^{N} P _i \pi_i (1 + (1+a_i)\delta _{i}+ a_i\delta_i^2),\\
&\text{subject to } \frac{1}{N}\sum_ {i=1}^{N} \pi_i(1+a_i \delta_i) \geqslant \ell.\\
\end{aligned}
\EQN\\
(\ref{eq:QPstlinear}) has a maximum if and only if its objective function is concave. However, this is satisfied when $a_i < 0$. 
Thus, we assume that $a_i \in (1-\frac{1}{\pi_i},0)$ for any $i \le N$.\\
In order to solve (\ref{eq:QPstlinear}), we use the quadratic programming method  summarised in Appendix A. 

\item Hereafter we shall assume that $b_i \neq 0$ implying that $\Psi_i$ is of the form
 $$\Psi_i:= \Psi_i(P_i,\delta_i)= \pi_i (1+a_i \delta_i+ b_i \delta_i^2).$$
We have that $\Psi_i \in (0,1)$ holds if and only if $a_i$ and $b_i$ satisfy the following conditions
\BQN\notag
\begin{cases}
a_i \in \Bigl(\text{max }(1-\frac{1}{\pi_i},-1-b_i), \text{min }(1+b_i,\frac{1}{\pi_i}-1)\Bigr),\\\notag
b_i \in (-1,0).
\end{cases}
\EQN
Moreover, (\ref{eq:optimprob}) can be rewritten as 
\BQN\label{eq:maxNLP}
\begin{aligned}
& \underset{\vk{\delta}}{\text{max }}  \sum_ {i=1}^{N} P _i \pi_i (1 + (1+a_i)\delta _{i}+ (a_i+b_i)\delta_i^2+b_i \delta_i^3),\\
&\text{subject to }\frac{1}{N}\sum_ {i=1}^{N} \pi_i(1+a_i \delta_i+b_i \delta_i^2) \geqslant \ell.\\
\end{aligned}
\EQN\\
Clearly, (\ref{eq:maxNLP}) is a non-linear optimisation problem subject to non-linear constraints. The most popular method discussed in the literature for solving this type of optimisation problem is the  {\it Sequential Quadratic Programming} method (SQP), see \cite{boggs1995sequential, nickel1984sequential,bartholomew2008nonlinear}.
It is an iterative method that generates a sequence of quadratic programs to be solved at each iterate. Typically, at a given iterate $x_k$, (\ref{eq:maxNLP}) is modelled by a QP subproblem subject to linear constraints and the solution to the latter is used as a search direction to construct a new iterate $x_{k+1}$. \COM{Henceforth, we are going to describe in more details the SQP algorithm.}\\
The optimisation problem at hand in this case  is given by
\BQN\label{eq:NLP}
\begin{aligned}
 &\underset{\vk{\delta}}{\text{min }}  f(\vk{\delta})= - \sum_ {i=1}^{N} P _i \pi_i (1 + (1+a_i)\delta _{i}+ (a_i+b_i)\delta_i^2+b_i \delta_i^3),\\
&\text{subject to }  g(\vk{\delta})= - \sum_ {i=1}^{N} \pi_i(1+a_i \delta_i+b_i \delta_i^2)+ N \ell \leq 0,
\end{aligned}
\EQN
where $f$ and $g$ are continuous and twice differentiable. See Appendix B for the main steps required to solve \eqref{eq:NLP}.\\

\subsubsection{Probability of renewal $\Psi_i$ as in \textbf{Mb)}}\label{probrenwMb}
We consider the case \textbf{Mb)} where the renewal probability is determined by the logistic regression model and is given by
\bqny{ 
\Psi_i:=\Psi_i(P_i, \delta_i)= \frac{1}{1+ c_i^{-1} e^{- T_i \delta_i }}, \quad 1 \le i \le N.
}
$c_i$ is a constant that depends on the probability of renewal  before premium change, $\pi_i$, and is given by
$$c_i=\frac{\pi_i}{1-\pi_i},$$	
and $T_i < 0$ is a constant (to be estimated in applications) that measures the elasticity of the policyholder relative to premium change.
The greater $|T_i|$, the more elastic the policyholder is to premium change.\\

In this regard, the optimisation problem can be formulated as follows
\BQN\label{eq:logit}
\begin{aligned}
& \underset{\vk{\delta}}{\text{max }}  \sum_ {i=1}^{N} P _i (1 + \delta _{i})/(1+e^{-T_i \delta_i}/c_i),\\
&\text{subject to } \frac{1}{N}\sum_ {i=1}^{N} 1/(1+e^{-T_i \delta_i}/{c_i}) \geqslant \ell.
\end{aligned}
\EQN	

(\ref{eq:logit}) is a non-linear optimisation problem subject to non-linear constraints. Therefore, in order to find the optimal solution $\vk{\delta}$, we use the SQP algorithm \mE{ described in Appendix B}.
\BRM\label{remark:logit}
It should be noted that \textbf{Mb)} can be approximated  by \textbf{Ma)} when the range of  $\delta_i$ is close to 0.
In this case, 
$$ \Psi_i(P_i,\delta_i)=\frac{c_i}{1+c_i}\Bigl(1+\frac{c_i T_i}{1+c_i} \delta_i-\frac{T_i^2 (c_i-1)}{2(1+c_i)^2} \delta_i^2\Bigr),$$
where
\BQN
\begin{aligned}
& \pi_i=\frac{c_i}{1+c_i}  ,
&  a_i=\frac{c_i T_i}{1+c_i}  \text{,      }
& \text{       }  b_i=-\frac{T_i^2 (c_i-1)}{2(1+c_i)^2}.
\end{aligned}
\EQN
\ERM
\subsubsection{Probability of renewal $\Psi_i$ as in \textbf{Mc)}}\label{subsubsection:Mc)}
Finally, we consider the case \textbf{Mc)} where \mE{the optimal solutions $\delta_i$ take its values from a discrete set. Also, the probabilities of renewal $\Psi_i$ at time 1 are fixed for each insured $i$ based on $\delta_i$ for  $i \leq N$}, as defined in Table \ref{table:example_Policy}. In this section, we deal with a {\it Mixed Discrete Non-Linear Programming} (MDNLP) optimisation problem. In this regard, we consider the discrete set $\mathbf{D}= \{-20\%,-15\%,-10\%,-5\%,0\%,5\%,10\%,15\%,20\%\}$ which corresponds to the optimal values that $\delta_i$ can take. Thus, (\ref{eq:optimprob})  can be reformulated as follows 

\BQN\label{eq:discreteopt}
\begin{aligned}
&\underset{\vk{\delta}}{ \text{min }} f(\vk{\delta})=- \sum_ {i=1}^{N} P _i (1 + \delta _{i,j}) \Psi_{i,j}(P_i,\delta_{i,j}),\\
&\text{subject to  } g(\vk{\delta})=- \sum_ {i=1}^{N} \Psi_{i,j}(P_i,\delta_{i,j}) + N  \ell  \leqslant  0   \text{              for     } j=1,\ldots,9, \\
& \text{and  }\vk{\delta} \in \mathbf{D}^N.
\end{aligned}
\EQN
In general, this type of optimisation problem is very difficult to solve due to the fact that the discrete space is non-convex. Several methods were discussed in the literature for the resolution of  (\ref{eq:discreteopt}), see \cite{arora1994methods}. \TE{The contribution} \cite {loh1991sequential} proposed a new method for solving the MDNLP optimisation problem subject to non-linear constraints. It consists \TE{in} approximating the original non-linear model by a sequence of mixed discrete linear problems evaluated at each  point iterate $\vk{\delta_k}$.  Also, a new method for solving a MDNLP was  introduced  by using a penalty function, see  the recent contribution \cite{MR3363684, ma2015exact} for more details. \mE{The algorithm for solving this type of optimisation problem is described in Appendix C.}
\\

\subsection {Maximisation of the retention level \atRR} \label{subSec:rent}
We consider the case where the insurer would like to  keep the maximum number of policyholders in the portfolio \atRR. Therefore, the optimisation problem of interest consists  \TE{in} finding the optimal retention level \atR whilst  increasing the expected premium volume by an amount say $C$ in the portfolio at time 1. 
Hence, the optimisation problem can be formulated as such
\BQN\label{eq:maxretention}
\begin{aligned}
&\underset{\vk{\delta}}{ \text{max }} \frac{1}{N} \sum_ {i=1}^{N} \Psi_i(P_i,\delta_i),\\
&\text{subject to } \mathbb{E}(P^{*})  \geqslant \mathbb{E}(P)+ C ,\\
\end{aligned}
\EQN
where\\
$\mathbb{E}(P^*)=\sum_ {i=1}^{N} P_i(1+\delta_i)\Psi_i(P_i,\delta_i)$ is the expected premium volume $@ \mathcal{R}$,    \\  
$\mathbb{E}(P)=\sum_ {i=1}^{N}P_i \pi_i$ is the expected premium volume at time 0,\\
and $C$ is a fixed amount which can be expressed as a percentage of the expected premium volume at time 0;
 it represents a loading to increase the premium volume at renewal.

In order to  solve (\ref{eq:maxretention}), we use the SQP algorithm described in Appendix B.
\COM{
 \section{optimisations}
 	Considering the logit model for the renewal probability in \eqref{eq: probRenewal}, our optimisation problem can be formulated as below 
 	 
 where 	$lb$ and $ub$ are the minimum and the maximum variation of premium, respectively  that can be set to the policyholders. It can be seen from \eqref{eq: obj const} that neither the objective function nor the constraints are convex. 
 
 \subsection{Algorithms}

The second approach consists in applying second order Taylor approximation on the objective function and the constraint function. This allows us to obtain a quadratic objective function with quadratic constraints by applying the second order Taylor approximation. The results of our approximations are summarised in the following equations
	\begin{itemize}
		\item \textbf{Taylor approximation of the objective function}
		\begin{equation}
		 \sum_ {i=1}^{n} P_0^i \left(
		 \frac{c_i}{ 1+c_i} +
		  \frac{c_i(1+c_i+T_i)}{(1+c_i)^2} \ \delta_i + 
		  \frac{2 c_i T_i + 2 c_i^2 T_i + c_i T_i^2 - c_i^2 T_i^2}{2(1+c_i)^3} \   \delta_i ^2 
		  + o[\delta_i^3] \right);
		\end{equation}	
		\item \textbf{Taylor approximation of the constraint function}
		\begin{equation}
		 \sum_ {i=1}^{n} 
		 \frac{c_i}{ 1+c_i} +
		  \frac{c_i(c_i T_i)}{(1+c_i)^2} \ \delta_i -
		  \frac{c_i T_i^2(c_i-1)}{2(1+c_i)^3} \   \delta_i ^2 
		  + o[\delta_i^3] \geqslant N.
		\end{equation}		
	\end{itemize}
	 To solve this quadratic problem, we have implemented the algorithm of [....] with the \textit{OPTI} toolbox of Matlab. With this algorithm, the optimisation problem is formulated in term of matrix as shown in the following equation 
	\footnote {We know that $Max \{f(\delta_i)\}=Min \{-f(\delta_i)\}$ so we can minimise the objective function $f(\delta_i)$ by multiplying it with $-1$.}: 
	\begin{equation}
		Min\ \frac{1}{2} \ \delta ^\top \ Q \delta \ + \ R^\top \ \delta,
	\end{equation}
	subject to	
	\begin{center}
		$\delta ^\top \ H \ \delta \ + \ \ K^\top \ \delta$,\\
		$lb \ \leqslant \ \delta \ \leqslant \ ub$,
	\end{center}	
where
	$\delta \ $ is a n-dimension vector  containing the premium change $\delta_i$ of each policyholder,\\
	$Q $ \footnote {We note that $Q$ is positive definite since the objective function is convex.}is a (n $\times$ n) matrix which consists of the quadratic coefficient of the objective function,\\
	$R$ is a n-dimension vector  containing the linear coefficient of the objective function,\\
	$H$\footnote {We note that $H$ is positive definite since the constraint function is convex.} is a (n $\times$ n) matrix which consists of the quadratic coefficient of the constraint function, \\
	$lb$ is a  n-dimension vector of the minimum value that can be chosen for $\delta$,\\
	$ub$ is a  n-dimension vector of the maximum value that can be chosen for $\delta$. 
Since our objective function and the constraint function are convex, the optimal solution of our problem is unique.
	\subsection{Case study}

\BIT
		\item the objective function and the constraints  used in the optimisation program are based on the objective and the constraints of the insurer respectively.   In particular, our optimisation will be based on  the expected profit and the expected volume (another risk measures can also be implemented). 
		\item Algorithm: since the corresponding  objective function and constraints   are non-convex, we implement the sequential quadratic program (SQP) approach. 
		\item case study: SQP, Taylor approximation (already solved with 1000 policyholders), simulations approach 
\EIT	
}
\section{Insurance Applications}
In this section, we consider a dataset that describes the production of the motor line of business of an insurance portfolio. We assume that the premiums are exponentially increasing. Also, the probability of renewal at time 0, $\pi_i$ for $i=1,\ldots,N$, are known and estimated by the insurance company for each category of policyholders based on historical data. Given that the behavior of the policyholders is unknown at the time of renewal, the probability of renewal at time 1, $\Psi_i$, depends on $\pi_i$ and $\delta_i$ for $i=1,\ldots,N$. If $\delta_i$ is positive, then $\Psi_i$ decreases whereas if $\delta_i$ is negative, it is more likely that the policyholder will renew his insurance policy  at time 1, thus generating a greater $\Psi_i$. In the following paragraphs, we are going to present some results relative to the optimisation problems at hand described in the last section.

\subsection{Optimisation problem Ma)} \label{subsec:Ma}
\subsubsection{Maximise the expected premium volume \atRR.} \label{subsubsec:MaxEPMa} 
We consider, first, the optimisation problem defined in (\ref{eq:QPstlinear}). In this case, the probability of renewal $\Psi_i$ is defined in \textbf{Ma)} and set $b_i=0$ for $i=1,\ldots,N$. 
Given that $a_i < 0 \text{ for } i\le N $, the probability of renewal $\Psi_i$ increases when $\delta_i$ is negative and decreases when $\delta_i$ is positive, thus describing perfectly the behavior of the policyholders that are subject to a decrease, respectively increase, in their premiums \atRR.
The table below describes some statistics on the data for $10'000$ policyholders.
\begin{center}
	\begin{tabular}{|c|c|}
		\hline
& Premium at time 0  \\

\hline
 \hline
Min  &200   \\
\hline 
Q1  &  491    \\
\hline 
Q2  &909  \\
\hline 
Q3  & 1'605    \\

\hline 
Max  & 9'061 \\
\hline
\hline
No. Obs. & 10'000   \\ 
\hline
Mean &   1'204  \\ 
\hline
Std. Dev. & 990  \\ 
\hline
		\end{tabular}
	\captionof{table}{Production statistics for the motor business. } \label{table: Premium Stat}
\end{center}

We consider that the insurance company would like to keep 85\% of its policyholders in its portfolio \atRR. By solving (\ref{eq:QPstlinear}) in Matlab with the function \textit{quadprog}, we  \TE{obtain} the optimal $\delta$ for each policyholder.

Next, we consider two scenarios:\\
\textbf{Scenario 1 }  The insurer would like to keep 75\% of the policyholders in his portfolio \atRR,\\
\textbf{Scenario 2 }   The insurer would like to keep 85\% of the policyholders in his portfolio \atRR. \\
Table \ref{table:scenario} below summarises the optimal results when solving (\ref{eq:QPstlinear}) and examines the effect of both scenarios on the expected premium volume and the expected  number of policyholders in the portfolio \atRR. \\ 
\begin{center}
	\begin{tabular}{|c||c|c||c|c|}
		\hline
	 Constraints on the retention level & \multicolumn{2}{|c||} {75\%}  & \multicolumn{2}{|c|} {85\%} \\
	 \hline
     Range of $\delta$ (\%) & (-10,20) & (-20,30)& (-10,20) & (-20,30)\\
     \hline
     Expected premium volume \atR (\%) &15.78 &23.03&8.70&12.96 \\
     \hline
     Expected number of policies \atR (\%)  &-3.52&-5.25&-0.16&-0.16\\
     \hline
     Average optimal delta (\%) & 19.99&29.90 &7.97 &11.89\\
     \hline
     Number of increases & 10'000 & 10'000& 6'196 &6'528\\
     \hline
     Number of decreases & - & - &3'804 &3'472 \\
     \hline
		\end{tabular}
	\captionof{table}{Scenarios testing.} \label{table:scenario}
\end{center}
\textbf{Scenario 1 } The optimal $\delta$ for both bounds corresponds approximately to the maximum value (upper bound) of the interval. This is mainly due to the fact that the insurer would like to keep only 75\% of the portfolio \atRR. Therefore, his main goal is to maximise the expected premium volume at time 1.\\
\textbf{Scenario 2 } For a retention level of 85\%, Table \ref{table:scenario} shows an increase in the expected premium volume which is less important than the one observed in Scenario 1. However, the expected number of policyholders in the portfolio \atR is higher and is approximately the same as at t=0.\\

Hereafter, we shall consider a retention level of  85\%. Usually, in practice, the size of a motor insurance portfolio exceeds $10'000$ policyholders. However, solving the optimisation problems for $\vk{\delta}$ using the described algorithms when $N$ is large requires a lot of time and heavy computation and may be costly for the insurance company. Thus, an idea to overcome this problem is to split the original portfolio into sub-portfolios and compute the optimal $\vk{\delta}$ for the sub-portfolios. 
One criteria that can be taken into account for the split is the amount of premium in our case. However, in practice, insurance companies have a more detailed data, thus  more information on each policyholders, so the criterion that are of interest for the split are the age of the policyholders, the car brand, car value \ldots 
Table \ref{table:split3} and Table \ref{table:split4} below describe the results when splitting the original portfolio into 3 and 4 sub-portfolios respectively.\\

\begin{center}
	\begin{tabular}{|c|c||c|c||}
		\hline
 \multicolumn{2}{|c||} {} &\multicolumn{2}{|c||} {Growth in \% \atR}  \\
	\hline
	Premium Range  &Average optimal $\delta$& Expected number of policies & Expected premium volume \\
	\hline
     $< 600$ &8.60\% & -0.27\% & 9.17\% \\
     \hline
     (600,1'200)  & 7.29\% & -0.03\% & 8.25\%\\
     \hline
     $> 1'200$  &8.05\% & -0.17\% & 8.99\% \\
     \hline
     After the split & 8.00\% & -0.16\% & 8.84\%\\
     \hline
    \hline
     Before the split  &7.97\% & -0.16\% & 8.70\% \\
     \hline 
     \hline
     Difference& - & 0\% & -0.13\% \\
     \hline
		\end{tabular}
	\captionof{table}{Split into 3 sub-portfolios.} \label{table:split3}
\end{center}	

\begin{center}
	\begin{tabular}{|c|c||c|c||}
		\hline
 \multicolumn{2}{|c||} {} &\multicolumn{2}{|c||} {Growth in \% \atR}  \\
	\hline
	Premium Range  &  Average optimal $\delta$ & Expected number of policies & Expected premium volume \\
	\hline
     $< 500$ & 8.99\% & -0.34\% & 9.50\% \\
     \hline
     (500, 800) & 6.27\%  & 0.15\% & 7.41\%\\
     \hline
     (800, 1'400) & 7.66\% &-0.09\% &8.49\%\\
     \hline
     $> 1'400$ &  8.47\% & -0.26\% & 9.31\% \\
     \hline
     After the split &  7.99\%  & -0.16\% & 8.95\%\\
     \hline
    \hline
     Before the split & 7.97\% & -0.16\% & 8.70\% \\
     \hline 
     \hline
     Difference &   & 0\% & -0.23\% \\
     \hline
		\end{tabular}
	\captionof{table}{Split into 4 sub-portfolios.} \label{table:split4}
\end{center}

In Table \ref{table:split3} and \ref{table:split4}, we consider that the insurer would like to keep 85\% of the policyholders in each sub-portfolios, thus a total of 85\% of the original portfolio. However, in practice, the constraints on the retention level \atR are specific to each sub-portfolio and this depending on the insurer's decision whether he would like to keep the policies with large premium amounts or small premium amounts in his portfolio \atRR.
In this regard, the insurance company sets the constraints on the expected number of policies for each sub-portfolios so that the constraint of the overall portfolio is approximately equal to 85\%. The error from the split into 3, respectively 4 sub-portfolios is relatively small and is of -0.13\%, respectively -0.23\% for the expected premium volume \atRR. It should be noted that the error margin increases as the number of splits increases.\\

\BRM
In the following sections, we limit the size of the insurance portfolio to 1'000 policyholders as the algorithms used thereafter to solve the optimisation problems are based on an iterative process and requires a lot of computation and time. Hence, an idea to solve the optimisation problem for an  insurance portfolio of size $n$ with $n\geq 1'000$ is to  split the original portfolio into sub-portfolios and compute the optimal results for the sub-portfolios, as discussed in the previous Section.
\ERM

	\subsubsection{Maximise the premium volume and minimise the variance of the  premium volume} \label{subsubsection:2objfunction}
		Similarly to the asset  allocation optimisation problem in finance \TE{ introduced by} Markowitz  \cite{markowitz1952portfolio}, the insurer performs a trade-off between the maximum aggregate expected premiums and the minimum variance of the total earned premiums; 
	\eEe{see also \cite{ValiRiskAllocation} for a different optimality criteria.}\\ 

We show next in
Table \ref{table: Volume and variance objective}  the optimal results  for the different constraints on the retention level and the possible range of premium changes. \\

\begin{center}
	\begin{tabular}{|c||c|c||c|c|}
		\hline
Retention level constraints    & \multicolumn{2}{|c||} {75\%}  & \multicolumn{2}{|c|} {85\%} \\
	 \hline
     Range of $\delta$ ( \%)  & (-10,20)  & (-20,30)& (-10,20) & (-20,30)\\
     \hline
     Aggregate expected future premiums \atR ( \%)  & 103.57 & 103.66 & 99.90
 & 103.90
\\
     \hline
  Variance of the aggregate future premiums  \atR ( \%)    & 109.76  &  113.08 & 98.41
 & 101.05
\\
       \hline
   Expected number of policies \atR ( \%)   & 98.95  & 98.84  & 99.98
  & 99.99
\\
     \hline
     Average optimal $\delta $ ( \%)  & 6.13  & 6.82  & 1.68
  & 1.98
\\
     \hline
       Average optimal increase  ( \%)  & 18.50 & 26.92  &  11.82
& 20.32
 \\
     \hline
         Average optimal decrease  ( \%)  & -8.33  & -16.32  & -8.92 & -17.10

 \\
     \hline
     Number of increases  & 539  & 535 & 511
 & 510\\
     \hline
     Number of decreases  & 461 & 465 & 489 & 490 \\
     \hline
		\end{tabular}
	\captionof{table}{Volume and variance objective optimal results.} \label{table: Volume and variance objective}
\end{center}
It can be seen that the optimal variance \atR increases with the range of the possible premium changes $\vk{\delta}$. For instance when  the insurer would like to keep $75\%$  of the policyholders,    the variance \atR increases from $109.76$ for $\delta \in (-10\%, 20\% ) $ to 	$113.08$ for  $\delta \in (-20\%, 30\% ) $, respectively. Furthermore, the increase in variance \atR is associated with an increase of the expected volume \atRR. This means that the riskier the portfolio the more the insurance company earns premiums. 
	
\subsubsection{Maximise the retention level \atRR.} We consider here that the insurer would like to maximise his retention level whilst increasing the expected premium volume \atR by a certain amount $C$ needed to cover, for instance, the operating costs and other expenses of the insurance company. $C$ can be expressed as a loading on the expected premium volume at time 0.  

In practice, the amount $C$ needed to cover the expenses of the company is set by the insurers. As stated previously, $C$ can be expressed as a percentage of the expected premium volume at time 0. Therefore, we consider three different loadings: 9\%, 10\% and 11\% thus adding an amount of $85'000$, respectively $95'000$ and $105'000$ to the expected premium volume at time 0. Also, we consider two ranges for $\delta$,   $\delta \in (-10\%,-20\%)$ and $\delta \in (-20\%,-30\%)$.
\COM{ Table \ref{table:scetestret} below shows the optimal results when decreasing, respectively increasing, the constraint amount $C$  to  for}

\begin{center}
	\begin{tabular}{|c||c|c||c|c||c|c|}
		\hline
 Constraint on the expected premium volume \atR  & \multicolumn{2}{|c||} {$C = 85'000$}  & \multicolumn{2}{|c||} {$C = 95'000$}  & \multicolumn{2}{|c|} {$C = 105'000$}\\
	 \hline
     Range of $\delta$ ( \%) & (-10,20)  & (-20,30)& (-10,20) & (-20,30)& (-10,20) & (-20,30)\\
     \hline
    Expected number of policies \atR ( \%) & -2.19    &-2.06   &-2.50  & -2.36  & -2.82& -2.67 \\
     \hline
Expected premium volume \atR ( \%) &  8.90   &8.90   &9.95  & 9.95 & 11.00& 11.00\\
\hline
     Average optimal $\delta $ ( \%) &13.92  &15.82  &15.12   &17.23  & 16.60 & 18.64\\
\hline
    
		\end{tabular}
	\captionof{table}{Scenario testing - Retention} \label{table:scetestret}
\end{center}
Table \ref{table:scetestret} shows that when $C$ increases, the expected number of policyholders \atR decreases whereas  the average optimal $\delta$ increases. 

\subsection{Optimisation problem Mb)} \label{subsec:Mb}  We consider  the optimisation problem defined in (\ref{eq:logit}) where the probability of renewal $\Psi_i$ is defined in \textbf{Mb)}. As discussed in Section \ref{probrenwMb}, $T_i$ describes the behavior of the policyholders subject to premium change. For instance, let's consider a policyholder whose probability of renewal without premium change $\pi_i$ is 0.95.  

In this Section, we will only consider the case where the insurer would like to maximise the expected premium volume \atRR. The constraint on the retention level is assumed to be of 85\%.

As stated in Section \ref{subsubsec:MaxEPMa},   insurers are more likely to increase the premiums of policyholders with small premium amounts and decrease the premiums of policyholders with large premium amounts. 

At the time of renewal, the insurer sets the constraints on the expected number of policyholders that he would like to keep in the portfolio. His decision is based on the maximum premium volume that he \TE{expects} to have \atRR. Typically, when the retention level is low, the expected premium volume \atR is greater compared to the case when the retention level is high. Therefore, we consider two scenarios:\\
\textbf{Scenario 1} The retention level  is of 75\%,\\
\textbf{Scenario 2} The retention level  is of 85\%.\\ 
The table below summarises the optimal results when solving (\ref{eq:logit}) for the different constraints.
\begin{center}
	\begin{tabular}{|c||c|c||c|c|}
		\hline
	 Constraints on the retention level & \multicolumn{2}{|c||} {75\%}  & \multicolumn{2}{|c|} {85\%} \\
	 \hline
     Range of $\delta$ (\%) & (-10,20) & (-20,30)& (-10,20) & (-20,30)\\
     \hline
     Expected premium volume \atR (\%) &17.84 &26.45&4.50&6.48 \\
     \hline
     Expected number of policies \atR (\%) &-0.93&-1.41&-0.02&-0.02\\
     \hline
     Average optimal delta (\%) & 20.00&30.00 &10.70 &16.09\\
     \hline
     Number of increases & 1'000 & 1'000& 703 &736\\
     \hline
     Number of decreases & - & - &297 &264 \\
     \hline
		\end{tabular}
	\captionof{table}{Scenarios testing.} \label{table:scenario_logitmodel}
\end{center}

\textbf{Scenario 1} Table \ref{table:scenario_logitmodel} shows that all policyholders are subject to an increase in their premiums and the average optimal $\delta$ for the whole portfolio corresponds to the maximum change in premium for both bounds of $\delta$. \\
\textbf{Scenario 2} As seen in Table \ref{table:scenario_logitmodel}, the expected number of policyholders \atR is approximately the same as the one before premium change. However, the growth in expected premium volume is lower than in Scenario 1 due to the fact that the average optimal $\delta$ for both bounds is lower.

\BRM
It should be noted that the probability of renewal defined in \textbf{Mb)} can be approximated by the probability of renewal defined in \textbf{Ma)} for $\delta$ relatively small (refer to Remark \ref{remark:logit}). 
Therefore, let's consider $\delta \in (-5\%,5\%)$ and a retention level $\ell = 85\%$ \atRR. The table below describes the optimal results when using the logit model \textbf{Mb)} and the polynomial model defined in \textbf{Ma)}.\\

\begin{center}
	\begin{tabular}{|c|c|c||c||}
		\hline
      Model & Logit & Polynomial & Difference\\
          \hline
     Growth in expected premium volume \atR &1.53\% &0.47\%&1.04\%\\
     \hline
     Growth in expected number of policies \atR &-0.02\%&-0.02\%&0\%\\
     \hline
     Average optimal delta & 2.97\%&1.30\% & - \\
     \hline
     Number of increases & 796 & 619& -\\
     \hline
     Number of decreases &204 & 381 & - \\
     \hline
		\end{tabular}
	\captionof{table}{Comparison between \textbf{Ma)} and \textbf{Mb)}.} \label{table:Difflogitmodelapp}
\end{center}

Table \ref{table:Difflogitmodelapp} shows that for a small range of $\delta$, the difference between the exact results obtained from \textbf{Mb)} and the approximate results obtained from \textbf{Ma)} is relatively small and is of around 1\% for the expected premium volume \atR and is of 0\% for the expected number of policyholders \atRR. Thus, the approximate values tend to the real ones when the range of $\delta$ tends to 0. 
 
\ERM
\subsection{Toy Models} In this Section, we consider two toy models. The first model consists in setting the same premium amounts among all policyholders.  Whereas in the second model, we assume that the policyholders have the same probability of renewal at time 0 irrespective of their premium amounts.\\
For both models, we compute the optimal results relative to the following scenarios:
\BIT
\item \textbf{Scenario 1}  Maximise the expected premium volume \atRR,
\item \textbf{Scenario 2}  Maximise the expected premium volume and minimise the corresponding variance \atRR,
\item \textbf{Scenario 3}  Maximise the retention level \atRR.\\
\EIT
\BIT
\item \textbf{Toy Model 1} The premiums are constant among all policyholders. We consider that $P_i=P= 200$ for all $i\leq N$.\\
\begin{center}
	\begin{tabular}{|c|c|c|c|}
		\hline
 Optimisation Problem & Scenario 1 & Scenario 2 & Scenario 3\\
\hline
    Constraints & 85\% & 85\%& C=20'000\\
     \hline
     Expected premium volume \atR (\%)&7.95 & 1.96  &11.76 \\
     \hline
     Expected number of policies \atR (\%)  &-0.02 & 0  &-1.34\\
     \hline
     Average optimal delta (\%) & 7.02 & 1.79  &12.55\\
     \hline
     		\end{tabular}
	\captionof{table}{Toy Model 1.} \label{table:toymodel1}
\end{center}

\item \textbf{Toy Model 2} The probabilities $\pi_i$ are constant at 0.9 for all policyholders.\\
\begin{center}
	\begin{tabular}{|c|c|c|c|}
		\hline
 Optimisation Problem  &Scenario 1 & Scenario 2  &  Scenario 3\\
\hline
    Constraints & 85\% & 85\%& C=20'000\\
 \hline
     Expected premium volume \atR (\%) &17.33 & 17.33 &7.78 \\
     \hline
     Expected number of policies \atR (\%)  &-2.22 & -0.73 &-0.03\\
     \hline
     Average optimal delta (\%) & 20.00 & 6.57 &0.24\\
     \hline
     		\end{tabular}
	\captionof{table}{Toy Model 2.} \label{table:toymodel2}
\end{center}
\EIT

These results are of interest when splitting the portfolios into sub-portfolios based on the premium amounts or the probability of renewal at time 0 of each policyholder.

\subsection{Optimisation problem Mc) and Simulation studies} \label{subsec:Mc} In this Section, we consider the case where the renewal probabilities $\Psi_i$ are fixed for each insured $i$, as defined in Table \ref{table:example_Policy}. To solve the optimisation problem (\ref{eq:optimprob}), we use the MDNLP method described in Appendix C. The table below summarises the optimal results for a portfolio of 100'000 policyholders with respect to different constraints on the retention level at renewal.\\
\COM{\begin{center}
	\begin{tabular}{|c|c|c|c|c|c|c|}
		\hline
    Retention level constraints (\%) & 85 & 87.5 & 90  & 92.5  & 95  & 97.5 \\
          \hline
     Growth in expected  premium volume \atR (\%) & 5.92 &5.92 &5.35 &4.18  &2.20    &-1.27  \\
     \hline
     Growth in expected number of policies \atR  (\%)&-7.89  &-7.89  &-5.26 &-2.63  &0.00   &2.63  \\
     \hline
     Average optimal delta (\%) &15.00 & 15.00 &10.00  &4.81 & -0.50 &-6.35  \\
     \hline
		\end{tabular}
	\captionof{table}{Scenario testing-Discrete optimisation } \label{table:DiscOpt}
\end{center}
For a portfolio of 100'000 policyholders}

\begin{center}
	\begin{tabular}{|c|c|c|c|c|c|c|}
		\hline
    Retention level constraints (\%) & 85 & 87.5 & 90  & 92.5  & 95  & 97.5 \\
          \hline
     Growth in expected  premium volume \atR (\%) & 5.92 &5.92 &5.34 &4.19  &2.22    &-1.24  \\
     \hline
     Growth in expected number of policies \atR  (\%)&-7.89  &-7.89  &-5.26 &-2.63  &0.00   &2.63  \\
     \hline
     Average optimal delta (\%) &15.00 & 15.00 &10.00  &4.82 & -0.51 &-6.37  \\
     \hline
		\end{tabular}
	\captionof{table}{Scenario testing-Discrete optimisation } \label{table:DiscOpt}
\end{center}
Table \ref{table:DiscOpt} shows that when the retention level increases, the expected number of policies increases whereas the expected premium volume \atR decreases. In fact, the average optimal $\vk{\delta}$ decreases gradually from 15\% for a retention level of 85\% to -6\% for a retention level of 97.5\%. Also, it can be seen that for a retention level of 95\%
 the optimisation has a negligible effect on the expected number of policies and premium volume \atR  as the average optimal $\vk{\delta}$ is approximately null. Hence, no optimisation is needed in this case. \\
In addition to the MDNLP approach, we have implemented a simulation technique which consists \TE{in} simulating the premium change  $\vk{\delta}$ for each policyholder as described in the following pseudo algorithm: \\
\BIT
\item \textbf{Step 1:}   Based on a chosen prior distribution for  $\vk{\delta}$,  sample the premium change for each policyholder,\\
\item \textbf{Step 2:}   Repeat \textbf{Step 1}   until the constraint on the retention level is satisfied, \\
\item \textbf{Step 3:}   Repeat \textbf{Step 2}     $m$ times,\\
\item \textbf{Step 4:}   Among the $m$ simulations  take the simulated $\vk{\delta}$ which gives out  the maximum expected profit.
\EIT
Next, we present the optimal results obtained through $1'000$ simulations for the same portfolio. We shall consider three different assumptions on the prior distribution of  $\vk{\delta}$, namely: 
\BIT
\item \textbf{ Case 1: Simulation based on the Uniform distribution} \\
In this simulation approach, we assume that the prior distribution of  $\vk{\delta}$ is uniformly distributed. As highlighted in Table \ref{table:param1}- \ref{table:param3}, the parameters of the uniform distribution  and the possible values of the premium change  are chosen so that the constraint on the retention level is fulfilled. We present in Table \ref{table:sim3} the simulation results.\\

\begin{center}
	\begin{tabular}{|c|c|c|c|c|c|c|}
		\hline
    Retention level constraints (\%) & 85 & 87.5 & 90  & 92.5  & 95  & 97.5 \\
          \hline
     Growth in expected  premium volume \atR (\%) & 5.13 &5.11 &4.02 &1.87 &-0.55    &-4.10 \\
     \hline
     Growth in expected number of policies \atR  (\% )&-10.32 &-6.64 &-5.24  &-2.61 & 0.04 &2.73   \\
     \hline
     Average optimal delta (\%) & 17.30&12.62  &9.95 &4.87 & -0.36 & -6.52\\
     \hline
		\end{tabular}
	\captionof{table}{Scenario testing- simulation approach: \TE{Uniform distribution.}} \label{table:sim3}
\end{center}

\item \textbf{ Case 2: Simulation based on practical experience }  \\ 
In this case we assume a prior distribution of $\vk{\delta}$ which is based on historical premium change  of each policyholder. 

\begin{center}
	\begin{tabular}{|c|c|c|c|c|c|c|}
		\hline
    Retention level constraints (\%) & 85 & 87.5 & 90  & 92.5  & 95  & 97.5 \\
          \hline
     Growth in expected  premium volume \atR (\%) & 5.50 &5.08 &4.10 &1.98 &-0.87    &-3.75 \\
     \hline
     Growth in expected number of policies \atR  (\% )&-9.19 &-7.70 &-5.26  &-2.63 & 0.47 &2.77   \\
     \hline
     Average optimal delta (\%) & 16.2&13.9  &10.0&4.92 & -1.17 & -6.25\\
     \hline
		\end{tabular}
	\captionof{table}{Scenario testing- simulation approach: \TE{practical experience}  } \label{table:sim2}
\end{center}

\COM{The tables below highlights the difference  between the simulated values and the real values.\\

\begin{center}
	\begin{tabular}{|c|c|c|c|c|c|c|}
		\hline
    Retention level constraints (\%) & 85 & 87.5 & 90  & 92.5  & 95  & 97.5 \\
          \hline
     Growth in expected  premium volume \atR (\%) & 0.74 &0.76 &1.26 &2.23 &2.72    &3.59  \\
     \hline
     Growth in expected number of policies \atR  (\% )&2.63 &-1.36 &-0.03  &-0.03 &-0.04 &-0.09   \\
     \hline
     Average optimal delta (\%) & -15.34&15.85  &0.51  &-0.89 & 29.8 & -12.4 \\
     \hline
		\end{tabular}
	\captionof{table}{Difference between the simulated values and the real ones } \label{table:sim}
\end{center}}

\item \textbf{Case 3: Simulation based on the results of  the MDNLP  }\\
We use   the distribution of  the optimal $\delta$ obtained from the MDNLP algorithm as a prior distribution. Table \ref{table:sim4} below summarises the optimal results. \\
\COM{\textbf{For 10'000 simulations, portfolio of 10'000 policyholders}\\

\begin{center}
	\begin{tabular}{|c|c|c|c|c|c|c|}
		\hline
    Retention level constraints (\%) & 85 & 87.5 & 90  & 92.5  & 95  & 97.5 \\
          \hline
     Growth in expected  premium volume \atR (\%) & 5.92 &5.92 &3.96 &1.68 &-0.83    &-3.94  \\
     \hline
     Growth in expected number of policies \atR  (\% )&-7.89 &-7.89 &-5.24  &-2.63 & 0.00 & 2.64   \\
     \hline
     Average optimal delta (\%) & 15.00&15.00  &9.96  &4.81 & -0.50 &-6.36  \\
     \hline
		\end{tabular}
	\captionof{table}{Scenario testing- simulation approach. } \label{table:sim4}
\end{center}

Difference between the simulated values and the real values\\

\begin{center}
	\begin{tabular}{|c|c|c|c|c|c|c|}
		\hline
    Retention level constraints (\%) & 85 & 87.5 & 90  & 92.5  & 95  & 97.5 \\
          \hline
     Growth in expected  premium volume \atR (\%) & 0.00 &0.00 &1.31 &2.40 &2.96    &2.70  \\
     \hline
     Growth in expected number of policies \atR  (\% )&0.00 &0.00 &-0.02  &0.00 &0.00 & -0.01   \\
     \hline
     Average optimal delta (\%) & 0.00&0.00  &0.38  &-0.05 & 0.89 &-0.17  \\
     \hline
		\end{tabular}
	\captionof{table}{Difference between the simulated values and the real ones } \label{table:sim}
\end{center}
\textbf{For 1'000 simulations, portfolio of 100'000 policyholders}\\}

\begin{center}\label{table:sim4}
	\begin{tabular}{|c|c|c|c|c|c|c|}
		\hline
    Retention level constraints (\%) & 85 & 87.5 & 90  & 92.5  & 95  & 97.5 \\
          \hline
     Growth in expected  premium volume \atR (\%) & 5.92 &5.92 &3.90 &1.61 &-0.91    &-4.05  \\
     \hline
     Growth in expected number of policies \atR  (\% )&-7.89 &-7.89 &-5.26  &-2.63 & 0.00 & 2.63  \\
     \hline
     Average optimal delta (\%) & 15.00&15.00  &10.00  &4.82 & -0.51 &-6.35 \\
     \hline
		\end{tabular}
	\captionof{table}{Scenario testing- simulation approach. } \label{table:sim}
\end{center}

\COM{The tables below highlights the difference  between the simulated values and the real values.\\

\begin{center}
	\begin{tabular}{|c|c|c|c|c|c|c|}
		\hline
    Retention level constraints (\%) & 85 & 87.5 & 90  & 92.5  & 95  & 97.5 \\
          \hline
     Growth in expected  premium volume \atR (\%) & 0.00 &0.00 &1.37 &2.48 &3.06    &2.84  \\
     \hline
     Growth in expected number of policies \atR  (\% )&0.00 &0.00 &0.00  &0.00 &0.00 & 0.00   \\
     \hline
     Average optimal delta (\%) & 0.00&0.00  &0.03  &0.01 & -0.27 & 0.21  \\
     \hline
		\end{tabular}
	\captionof{table}{Difference between the simulated values and the real ones } \label{table:sim}
\end{center}}

\COM{\item \textbf{Distribution based simulation:} In this case, we first generate random numbers in (0,1) from the uniform distribution. Then,  we find the optimal $\delta$  for each policyholder by using the inversion method and this, based on the empirical quantile distribution  of the real optimal $\delta$ obtained from the MDNLP algorithm. The table below describes the results of the simulation.\\
\textbf{For 1'000 simulations, portfolio of 10'000 policyholders}\\

\begin{center}
	\begin{tabular}{|c|c|c|c|c|c|c|}
		\hline
    Retention level constraints (\%) & 85 & 87.5 & 90  & 92.5  & 95  & 97.5 \\
          \hline
     Growth in expected  premium volume \atR (\%) & 5.92 &5.92 &3.95 &1.66 &-0.85    &-3.98  \\
     \hline
     Growth in expected number of policies \atR  (\% )&-7.89 &-7.89 &-5.26  &-2.61 & 0.00 & 2.63   \\
     \hline
     Average optimal delta (\%) & 15.00&15.00  &10.00  &4.78 & -0.51 &-6.32  \\
     \hline
		\end{tabular}
	\captionof{table}{Scenario testing- simulation approach } \label{table:sim}
\end{center}

\COM{Difference between the simulated values and the real values\\

\begin{center}
	\begin{tabular}{|c|c|c|c|c|c|c|}
		\hline
    Retention level constraints (\%) & 85 & 87.5 & 90  & 92.5  & 95  & 97.5 \\
          \hline
     Growth in expected  premium volume \atR (\%) & 0.00 &0.00 &1.33 &2.42 &2.98    &2.74  \\
     \hline
     Growth in expected number of policies \atR  (\% )&0.00 &0.00 &0.00  &-0.02 &0.00 & 0.00   \\
     \hline
     Average optimal delta (\%) & 0.00&0.00  &0.05  &0.74 & -0.30 &0.50  \\
     \hline
		\end{tabular}
	\captionof{table}{Difference between the simulated values and the real ones. } \label{table:sim}
\end{center}
\textbf{ For 10'000 simulations, portfolio of 10'000 policyholders}\\
\begin{center}
	\begin{tabular}{|c|c|c|c|c|c|c|}
		\hline
    Retention level constraints (\%) & 85 & 87.5 & 90  & 92.5  & 95  & 97.5 \\
          \hline
     Growth in expected  premium volume \atR (\%) &5.92  &5.92 &3.96 &1.67  &-0.79   &-3.95  \\
     \hline
     Growth in expected number of policies \atR  (\% )&-7.89 &-7.89   & -5.25&-2.63& 0.00&2.64 \\
     \hline
     Average optimal delta (\%) &15.00 & 15.00 &9.97   &4.80  & -0.50&-6.35\\
     \hline
		\end{tabular}
	\captionof{table}{Scenario testing- simulation approach } \label{table:sim}
\end{center}}

Difference between the simulated values and the real values\\

\begin{center}
	\begin{tabular}{|c|c|c|c|c|c|c|}
		\hline
    Retention level constraints (\%) & 85 & 87.5 & 90  & 92.5  & 95  & 97.5 \\
          \hline
     Growth in expected  premium volume \atR (\%) & 0.00 &0.00 &1.32 &2.41 &2.92    &2.72  \\
     \hline
     Growth in expected number of policies \atR  (\% )&0.00 &0.00 &-0.02  &0.00 &0.00 & -0.01   \\
     \hline
     Average optimal delta (\%) & 0.00&0.00  &0.33  &0.17 & 1.09 &-0.07  \\
     \hline
		\end{tabular}
	\captionof{table}{Difference between the simulated values and the real ones} \label{table:sim}
\end{center}

\textbf{For 1'000 simulations, portfolio of 100'000 policyholders}\\

\begin{center}
	\begin{tabular}{|c|c|c|c|c|c|c|}
		\hline
    Retention level constraints (\%) & 85 & 87.5 & 90  & 92.5  & 95  & 97.5 \\
          \hline
     Growth in expected  premium volume \atR (\%) & 5.92 &5.92 &3.90 &1.60 &-0.91    &-4.06  \\
     \hline
     Growth in expected number of policies \atR  (\% )&-7.89 &-7.89 &-5.26  &-2.62 & 0.00 & 2.63   \\
     \hline
     Average optimal delta (\%) & 15.00&15.00  &10.00  &4.81 & -0.51 &-6.36  \\
     \hline
		\end{tabular}
	\captionof{table}{Scenario testing- simulation approach } \label{table:sim}
\end{center}

Difference between the simulated values and the real values\\

\begin{center}
	\begin{tabular}{|c|c|c|c|c|c|c|}
		\hline
    Retention level constraints (\%) & 85 & 87.5 & 90  & 92.5  & 95  & 97.5 \\
          \hline
     Growth in expected  premium volume \atR (\%) & 0.00 &0.00 &1.37 &2.49 &3.07    &2.85  \\
     \hline
     Growth in expected number of policies \atR  (\% )&0.00 &0.00 &0.00  &-0.01 &0.00 & 0.00   \\
     \hline
     Average optimal delta (\%) & 0.00&0.00  &0.02  &0.28 & -1.24 &0.16  \\
     \hline
		\end{tabular}
	\captionof{table}{Difference between the simulated values and the real ones } \label{table:sim}
\end{center}}
\EIT
It can be seen that the simulation approaches yield  approximately to the same results as the MDNLP algorithm \TE{presented in Table \ref{table:DiscOpt}.}

\section{Appendix A: Constrained quadratic programming}
We present next the steps for the quadratic programming method utilised in our paper. 

\textbf{Step 1:} (\ref{eq:QPstlinear}) can be reformulated as follows

\BQN\label{eq:QPstlinearmat}
\begin{aligned}
& \underset{\vk{\delta}}{ \text{min }} f(\vk{\delta})= \frac{1}{2} \vk{\delta}^{\top} Q \vk{\delta} + \vk{c}^\top \vk{\delta},\\
& \text{subject to } g(\vk{\delta}) =  \vk{A}^\top  \vk{\delta} \le   b , 
\end{aligned}
\EQN
where $\vk{\delta}=(\delta_1,\ldots,\delta_N)^\top$ and $\vk{c}$ is a vector describing the coefficient of the linear terms of $f$ given by
$$\vk{c}= (-\pi_1 P_1(1+a_1),\ldots,-\pi_N P_N(1+a_N))^\top.$$
Here $Q$ is a diagonal and positive definite matrix describing the coefficients of the quadratic terms of $f$ determined by 
$$Q= \begin{pmatrix} 
 -2\pi_1P_ {1}a_1 &0&0 &\ldots\ & 0  \\
 0& -2\pi_2 P_ {2}a_2&0 &\ldots &0\\
0&\ldots&-2\pi_iP_ {i}a_i &\ldots&0\\
 0&0&0 &\ldots &-2\pi_NP_ {N}a_N\\
 \end{pmatrix}.$$\\
Since (\ref{eq:QPstlinearmat}) has only one constraint, $\vk{A}$ is a vector related to the linear coefficients of $g$ given by $$\vk{A}= - ( \pi_ {1}a_1 ,\pi_ {2}a_2,\ldots,\pi_Na_N)^\top,$$
and finally,  $b= N \ell - \sum_{i=1}^{N} \pi_i$.\\
It should be noted that the constant term of the objective function $f$ is not accounted for in the resolution of (\ref{eq:QPstlinearmat}). \\
\textbf{Step 2:} Let $\mathcal{L}(\vk{\delta},{\lambda})=f(\vk{\delta})+{\lambda} g(\vk{\delta})$ be the Lagrangian function of  (\ref{eq:QPstlinearmat}) where $\lambda$ is the Lagrangian multiplier. \\
Given that $Q$ is a positive definite matrix, the well-known Karush-Kuhn-Tucker (KKT) conditions (\TE{see for details \cite{MR3363684}[page 342]}) defined below are  sufficient for a global minimum of (\ref{eq:QPstlinearmat}) if they are satisfied for a given vector  $(\vk{\delta}^*,{\lambda}^*)$
\BQN \label{Pr1} \left\{
			 	\begin{array}{lcl}
         			\nabla \mathcal{L}(\vk{\delta}^*,\lambda^*)&=& 0,\\
         			{{\lambda}^{*}} g(\vk{\delta}^*) &=& 0, \\
         			g(\vk{\delta}^*) & \le & 0, \\
         			\lambda^* &\ge & 0.
              	\end{array}
      \right.
\EQN
\BRM
In the settings of \textbf{Problem T1a} where the insurer aims at maximising the premium volume and  minimising   the corresponding variance, the optimisation problem can be expressed as follows 
\BQN\label{eq:doubleObj}
\begin{aligned} 
&\underset{\vk{\delta}}{\text{max }} \sum_ {i=1}^{N} P _i \pi_i (1 + (1+a_i)\delta _{i}+ a_i\delta_i^2),  \\
&
\underset{\vk{\delta}}{\text{min }} \sum_ {i=1}^{N} P _i \pi_i (1 + (1+a_i)\delta _{i}+ a_i\delta_i^2) (1- \pi_i (1+a_i\delta _{i})),\quad \\
& \text{subject to  } \frac{1}{N} \sum_ {i=1}^{N} \pi_i(1+a_i \delta_i) \geqslant \ell.\\
\end{aligned}
\EQN\\
\ERM

\section{Appendix B:  Solution of \eqref{eq:NLP}}
\textbf{ Step 1:} Let $\mathcal{L}(\vk{\delta},\lambda)=f(\vk{\delta})+\lambda g(\vk{\delta})$ be the Lagrangian function of  (\ref{eq:NLP}) where $\lambda \in \R$ is the Lagrangian multiplier and ($\vk{\delta_0}$,$\lambda_0$) an initial estimate of the solution. It should be noted that the SQP is not a feasible point method. This means that neither the initial point nor the subsequent iterate ought to satisfy the constraints of the optimisation problem. \\

\textbf{Step 2:}  In order to find the next point iterate $(\vk{\delta_1},\lambda_1)$, the SQP determines a step vector $\vk{s}=(\vk{s_\delta},s_\lambda)$  solution of the QP sub-problem evaluated  at ($\vk{\delta_0}$,$\lambda_0$) and  defined below
\begin{equation}\label{eq:subQP}
\begin{aligned}
&\underset{\vk{s}}{\text{min}}
&&\frac{1}{2}\vk{s}^\top  H \vk{s} +\nabla f(\vk{\delta_0})^\top \mathbf{\vk{s}},\\
&\text{subject to }
&& \nabla g(\vk{\delta_0})^\top \vk{s}+ g(\vk{\delta_0})\leq 0,\\
\end{aligned}
\end{equation}
where $H$ is an approximation of the Hessian matrix of $\mathcal{L}$, $\nabla f$ and $\nabla g$ are the gradient of the objective and the constraint functions respectively.\\
The Hessian matrix $H$ is updated at each iteration by the BFGS quazi Newton formula. The SQP method maintains the sparsity of the approximation of the Hessian matrix and its positive definetness, a necessary condition for a unique solution.\\

\textbf{Step 3:}  In order to ensure the convergence of the SQP method to a global solution,  the latter uses a merit function $\phi$ whose reduction implies progress towards a solution. Thus, a step length, denoted by $\alpha \in (0,1)$,  is chosen in order to guarantee the reduction of $\phi$ after each iteration such that 
$$\phi (\delta_k+\alpha s_k) \leq \phi(\delta_k),$$
with $$ \phi (x)= f(x)+r g(x)  \text{  and  }   r>|\lambda|.$$\\

\textbf{Step 4:} The new point iterate is given by  $(\vk{\delta_1},\lambda_1)=(\vk{\delta_0}+\alpha \vk{s_{\delta}}, \lambda_0+\alpha s_{\lambda})$. If $(\vk{\delta_1},\lambda_1)$ satisfies the KKT conditions (\ref{Pr1}), the SQP converges at that point.
If not, set $k=k+1$ and go back to \textbf {Step 2}.
\EIT

\BRM 
It should be noted that  the KKT conditions defined in (\ref{Pr1}) are known as the first order optimality conditions, \TE{see e.g., \cite{MR3363684}}. Hence, if, for a given vector $(\vk{\delta^*},\lambda^*)$, the KKT conditions are satisfied, then $(\vk{\delta^*},\lambda^*)$ is a local minimum of (\ref{eq:NLP}).
\ERM

\section{Appendix C: MDNLP optimisation problem \eqref{eq:discreteopt} }
\textbf{Step 1:} Given that $\Psi_i$ is discrete and depends on the values of $\delta_i$, we assume that $\Psi_i$ can be written as a function of $\delta_i$ as follows
$$ \Psi_i(\delta_i)= -0.9775 \delta_i^2-0.4287 \delta_i +0.9534     \text{     for } \delta_i \in \mathbf{D}$$
(\ref{eq:discreteopt})  is then treated  as a continuous optimisation problem and the optimal solution is found by using one of the methods described previously.   We denote by $\vk{\delta^*}$ the continuous optimal solution. \\

\textbf{Step 2:} Let $\vk{\delta_0}$ be the rounded up vector of $\vk{\delta^*}$ to the nearby discrete values of the set $\mathbf{D}$. $\vk{\delta_0}$ is considered to be the initial point iterate. If $\vk{\delta_0}$ is not a feasible point of (\ref{eq:discreteopt}), then (\ref{eq:discreteopt}) is approximated by a mixed discrete linear optimisation problem at $\vk{\delta_0}$  and is given by 
\BQN\label{eq:MDLP}
\begin{aligned}
&\underset{\vk{\delta}}{ \text{min }} \nabla f(\vk{\delta_0})^\top  (\vk{\delta}-\vk{\delta_0}),\\
&\text{subject to  } g(\vk{\delta_0})+\nabla g(\vk{\delta_0})^\top (\vk{\delta}-\vk{\delta_0})\leqslant  0,\\
& \text { and }\vk{\delta} \in \mathbf{D}^N.
\end{aligned}
\EQN

\textbf{Step 3:} (\ref{eq:MDLP}) is solved by using a linear programming method and the branch and bound method, see \cite{dakin1965tree} for more details. We denote by $\vk{\delta_k}$ the new point iterate. If $\vk{\delta_k}$ is feasible and $||\vk{\delta_k}-\vk{\delta_{k-1}}|| < \epsilon$ with $\epsilon>0$ small, then the iteration is stopped.
Else $k=k+1$ and go back to \textbf{Step 2}.
\BRM
If, for a certain point iterate $\vk{\delta}$, the constraint of (\ref{eq:discreteopt}) is satisfied and $\vk{\delta} \in \mathbf{D}^N$   then $\vk{\delta}$ is a feasible solution of the optimisation problem.\\
In general, it is very hard to find the global minimum of a MDNLP optimisation problem due to the fact that there are multiple local minimums.
Therefore, $\vk{\delta^*}$ is said to be a global minimum if  $\vk{\delta^*}$  is feasible and $f(\vk{\delta^*})\leq f(\vk{\delta})$ for all feasible $\vk{\delta}$.
\ERM
\section{Appendix D: Prior distribution for simulation } \label{appendix: Prior}
\subsection{\TE{Simulation based on the Uniform distribution \TE{(simulation Case 1)}}} \TE{The tables below describe the range of $\delta$  with their respective distribution based on the different retention levels.}

\begin{center}
	\begin{tabular}{|c|c|c|c|c|}
		\hline
   Retention level (\%) & 85 & 87.5 & 90     \\
          \hline
     Range of $\vk{\delta} (\%)$ & $\{15, 20\}$   & $\{10, 15\}$ & $ \{0,5,10, 15\}$   \\
     \hline
      Prior distribution & $U(0.85,0.99)$  & $U(0.90,0.99)$  & $U(0.04,0.68)$     \\
     \hline
\end{tabular}
	\captionof{table}{Possible range of $\vk{\delta}$ and prior distribution uniformly distributed.} \label{table:param1}
\end{center}

\begin{center}
	\begin{tabular}{|c|c|c|c|}
		\hline
   Retention level (\%) & 92.5 & 95 & 97.5     \\
          \hline
      Range of $\vk{\delta} (\%)$ & $ \{-5,0,5,10, 15\}$   &  $ \{-5,0,5,10, 15\}$ &  $ \{-20,-10,-5,0,5,10, 15\}$    \\
     \hline
      Prior distribution & $U(0.05,0.40)$  & $U(0.04,0.21)$ & $U(0.002,0.47)$    \\
     \hline
\end{tabular}
	\captionof{table}{Possible range of $\vk{\delta}$ and prior distribution uniformly distributed.} \label{table:param3}
\end{center}



\bibliographystyle{ieeetr}

\bibliography{priceoptG}
	
\end{document}